\documentclass[twocolumn,showpacs,preprintnumbers,amsmath,amssymb]{revtex4}

\usepackage{graphicx}
\usepackage{dcolumn}
\usepackage{bm}

\begin{document}

\preprint{APS/123-QED}

\title{Decoherence induced by squeezing control errors in optical and ion trap holonomic quantum
computations}

\author{Viatcheslav I. Kuvshinov}
\email{v.kuvshinov@sosny.bas-net.by}
\affiliation{%
Joint Institute for Power and Nuclear Research, 220109 Krasina
str., 99, Minsk, Belarus. }

\author{Andrei V. Kuzmin}
\email{avkuzmin@sosny.bas-net.by}
\affiliation{%
Joint Institute for Power and Nuclear Research, 220109 Krasina
str., 99, Minsk, Belarus.
}

\date{\today}

\begin{abstract}
We study decoherence induced by stochastic squeezing control
errors considering the particular implementation of Hadamard gate
on optical and ion trap holonomic quantum computers. We find the
fidelity for Hadamard gate and compute the purity of the final
state when the control noise is modeled by Ornstein-Uhlenbeck
stochastic process. We demonstrate that in contradiction to the
case of the systematic control errors the stochastic ones lead to
decoherence of the final state. In the small errors limit we
derive a simple formulae connecting the gate fidelity and the
purity of the final state.
\end{abstract}

\maketitle

Holonomic quantum computations exploiting non-abelian geometrical
phases~\cite{Wilczek1} was primarily proposed in the
Ref.~\cite{HQC1} and developed further in the Ref.~\cite{HQC2}.
Many implementations of holonomic quantum computers (HQC) have
been proposed. Particularly, the realization of HQC within quantum
optics was suggested (optical HQC)~\cite{OHQC}. Laser beams in a
non-linear Kerr medium were exploited for this purpose. Two
different sets of control devices can be used in this case. The
first one considered in this Rapid Communication consists of one-
and two-mode displacing and squeezing devices. The second one
includes SU(2) interferometers. As well trapped ions with the
excited state connected to a triple degenerate subspace (four
level $\Lambda$-system) can be used to implement HQC~\cite{trio}.
Another approach to HQC exploiting squeezing and displacement of
the trapped ions vibrational modes was suggested in the
Ref.~\cite{TRIOHQC}. This implementation of HQC is mathematically
similar to the first embodiment of the optical HQC~\cite{OHQC} and
thus it is also considered in this work. Particularly, expressions
for the adiabatic connection and holonomies are the same in these
cases. Another proposed implementation of HQC was the HQC with
neutral atoms in cavity QED~\cite{cavity}. The coding space was
spanned by the dark states of the atom trapped in a cavity.
Dynamics of the atom was governed by the generalized
$\Lambda$-system Hamiltonian. Mathematically similar
semiconductor-based implementation of HQC was proposed in the
Refs.~\cite{semicond}, where one-qubit gates were also realized in
the framework of the generalized $\Lambda$-system. In distinction
from the cavity model of HQC its physical implementation exploits
semiconductor excitons driven by sequences of laser
pulses~\cite{semicond}. For the two-qubit gate implementation the
bi-excitonic shift was used. The generalized $\Lambda$-system with
the different Rabi frequencies parametrization was exploited
recently for HQC implemented by Rf-SQUIDs coupled through a
microwave cavity~\cite{SQUID}. One more solid state implementation
of HQC based on Stark effect was proposed in the
Ref.~\cite{Stark}.

Let us briefly remind the main results concerning the holonomic
quantum computation. In HQC non-abelian geometric phases
(holonomies) are exploited to implement unitary transformations
over the quantum code. The later is some degenerate subspace $C^N$
spanned on eigenvectors of Hamiltonian $H_0$, which initiates the
parametric isospectral family of Hamiltonians $F = \{ H(\lambda) =
U(\lambda) H_0 U^\dag (\lambda)\}_{\lambda \in M}$. Here
$U(\lambda)$ is a unitary operator, $\lambda$ is a vector
belonging to the space of the control parameters $M$ and $N$
denotes the dimension of the degenerate computational
subspace~\cite{HQC1,HQC2}. Quantum gates are implemented when the
control parameters are adiabatically driven along the loops in the
control manifold $M$. The unitary operator mapping the initial
state vector belonging to $C^N$ into the final one has the form
$e^{i \phi} \Gamma_\gamma (A_\mu)$, where the index $\mu$
enumerates control parameters $\lambda_\mu$ constituting vector
$\lambda$ and $\phi$ is the dynamical phase. Holonomy associated
with the loop $\gamma \in M$ is
\begin{equation}\label{holonomy}
    \Gamma_\gamma (A_\mu) = {\bf \hat{P}} \exp{\left\{ \int\limits_\gamma A_\mu d \lambda_\mu
    \right\}}.
\end{equation}
Here ${\bf \hat{P}}$ denotes the path ordering operator, $A_\mu$
is the matrix valued adiabatic connection given by the
expression~\cite{Wilczek1}:
\begin{equation}\label{connection}
    \left( A_\mu \right)_{mn} = \left< \varphi_m \right| U^\dag
    \frac{\partial}{\partial \lambda_\mu} U \left| \varphi_n
    \right> ,
\end{equation}
where $\left| \varphi_k \right>$ with $k = \overline{1,N}$ are the
eigenvectors of the Hamiltonian $H_0$ forming the basis in $C^N$.
Dynamical phase $\phi$ will be omitted bellow due to the suitable
choice of the zero energy level. We shall consider the single
subspace $C^N$ (no energy level crossings are assumed).

It is evident that the quantum gate (holonomy) performed depends
on the path passed in the control parameters space. As well it is
obvious that in real experiments it is impossible to pass the
desired loop in the control manifold without any deviations.
Errors in the assignment of the classical control parameters
$\lambda$ are unavoidable. The question about robustness of
holonomic quantum computations with respect to the control errors
has attracted a lot of attention recently. Namely, the effect of
the errors originated from the imperfect control of classical
parameters was studied for ${\bf CP}^n$ model of HQC in the
Ref.~\cite{Err1} where the control-not and Hadamard gates were
particularly considered. Berry phase for the spin $1/2$ particle
in a classical fluctuating magnetic field was considered in the
Ref.~\cite{spin12}. Approach based on the non-abelian Stokes
theorem~\cite{Stokes} was proposed in the Ref.~\cite{wePLA1}.
Namely, the general expression for the fidelity valid for
arbitrary implementation of HQC in the case of the single
systematic control error having arbitrary size and duration was
derived. Simple approximate formulae was found in the small error
limit. Adiabatic dynamics of quantum system coupled to a noisy
classical control field was studied in the Ref.~\cite{Gaitan}. It
was demonstrated that stochastic phase shift arising in the
off-diagonal elements of the system's density matrix can cause
decoherence. The efficiency of Shor algorithm~\cite{Shor1} run on
a geometric quantum computer was investigated in the case when the
decoherence induced by the stochastic control errors was taken
into account. The study of the robustness of the non-abelian
holonomic quantum gates with respect to the stochastic
fluctuations of the control parameters was presented in the
Ref.~\cite{ThreeRegimes}. Three stability regimes were
discriminated in this work for the HQC model with qubits given by
polarized excitonic states controlled by laser pulses. Noise
cancellation effect for simple quantum systems was considered in
the Ref.~\cite{Solinas04}. Robustness of the parametric family of
quantum gates subjected to stochastic fluctuations of the control
parameters was studied in the Ref.~\cite{Zhu04}. Usage of the
cyclic states~\cite{cyclicSt} allowed to consider quantum gates
which could be continuously changed from dynamic gates to purely
geometric ones. It was shown that the maximum of the gate fidelity
corresponds to quantum gates with a vanishing dynamical phase.
Robust Hadamard gate implementation for optical~\cite{OHQC} and
ion trap~\cite{TRIOHQC} holonomic quantum computers was proposed
in the Ref.~\cite{wePLA2}. The cancellation of the small squeezing
control errors up to the fourth order on their magnitude was
demonstrated. Hadamard gate is one of the key elements of the main
quantum algorithms, for instance see~\cite{Shor1, Grover1}. Thus
the search for its robust implementations is of importance.

During the last few years much attention has been payed to the
study of both abelian and non-abelian geometric phases in the
presence of decoherence which is the most important limiting
factor for quantum computations. Let us briefly overview some of
these works. The abelian geometric phase of the two-level quantum
system interacting with a one and two mode quantum field subjected
to the decoherence was considered in the Ref.~\cite{DecohField}.
It was demonstrated that when the geometric phase is generated by
an adiabatic evolution the first correction due to the decoherence
of the driving quantized field for the no-jump trajectory has the
second order in the decaying rate of the field but it is not the
case for the non-adiabatic evolution. Non-abelian holonomies in
the presence of decoherence were investigated in the
Ref.~\cite{Guridi} using the quantum jump approach. The effects of
environment on a universal set of holonomic quantum gates were
analyzed. Refocusing schemes for holonomic quantum computation in
the presence of dissipation were discussed in the
Ref.~\cite{refocus}. It has been shown that non-abelian geometric
gates realized by means of refocused double-loop scheme possessed
a certain resilience against decoherence. Quantum Langevin
approach has been used to study the evolution of two-level system
with a slowly varying Hamiltonian and interacting with a quantum
environment modeled as a bath of harmonic
oscillators~\cite{HarBath}. It allowed to obtain the dissipation
time and the correction to Berry phase in the case of adiabatic
cyclic evolution. The realization of universal set of holonomic
quantum gates acting on decoherence-free subspaces has been
proposed in the Ref.~\cite{DecohFree}. It has been shown how it
can be implemented in the contexts of trapped ions and quantum
dots. The performance of holonomic quantum gates in semi-conductor
quantum dots under the effect of dissipative environment has been
studied in the Ref.~\cite{Dissipation}. It was demonstrated the
influence of the environment modeled by the superhomic thermal
bath of harmonic oscillators could be practically suppressed. The
study of the non-adiabatic dynamics and effects of quantum noise
for the ion trap setup proposed in the Ref.~\cite{trio} has been
also done~\cite{ref29}. The optimal finite operation time was
determined. In the references mentioned above the fidelity was
used as the main measure of gate resilience.

In this Rapid Communication we consider optical and ion trap
implementations of HQC proposed in the Refs.~\cite{OHQC}
and~\cite{TRIOHQC} respectively. Regarding the particular
implementation of Hadamard gate we study the decoherence induced
by stochastic squeezing control errors. Following the
Ref.~\cite{spin12} we model the random fluctuations by
Ornstein-Uhlenbeck stochastic process. We analytically obtain the
gate fidelity and the final state purity as the measures of the
gate robustness with respect to the decoherence induced by
stochastic control errors. In the small squeezing control errors
limit we derive a simple formulae connecting the gate fidelity and
the purity of the final state. As well we demonstrate that
systematic control errors do not lead to the decoherence. The
systematic error means the error equal for all qubits in the
ensemble or for all consecutive gate implementations performed on
the given qubit.

One-qubit gates are given as sequence of single mode squeezing and
displacing operations~\cite{OHQC, TRIOHQC}:
\begin{equation}\label{UDS}
    U (\eta, \nu) = D(\eta) S(\nu),
\end{equation}
where
\begin{eqnarray}\label{SD}
    &&S(\nu) = \exp{\left( \nu a^{\dag 2} - \bar{\nu} a^2
    \right)}, \nonumber \\
    &&D(\eta) = \exp{\left( \eta a^\dag - \bar{\eta} a \right)}
\end{eqnarray}
denote single mode squeezing and displacing operators
respectively, $\nu = r_1 e^{i \theta_1}$ and $\eta = x +iy$ are
corresponding complex control parameters, $a$ and $a^\dag$ are
annihilation and creation operators. The line over the parameter
denotes complex conjugation. The expressions for the adiabatic
connection and the curvature tensor can be found in the
Refs.~\cite{OHQC, TRIOHQC}. Following our previous
Letter~\cite{wePLA2} we consider Hadamard gate
\begin{equation}\label{Hadamard0}
    H_0 = \frac{1}{\sqrt{2}} \left(%
\begin{array}{cc}
  1 & 1 \\
  1 & -1 \\
\end{array}%
\right)
\end{equation}
implemented when two rectangular loops belonging to the planes
$(x,r_1)\left|_{\theta_1 = 0} \right.$ and
$(y,r_1)\left|_{\theta_1 = 0} \right.$ are passed. Namely,
\begin{equation}\label{HGG}
    -iH_0 = \Gamma (C_{II})\left|_{\Sigma_{II} = \pi /2} \right. \Gamma
    (C_I) \left|_{\Sigma_I = \pi /4} \right. ,
\end{equation}
where the holonomies are
\begin{eqnarray}\label{Gammas}
    \Gamma (C_I) = \exp{\left( -i\sigma_y \Sigma_I \right)}, \quad
    \Sigma_I = \int\limits_{S(C_I)} dx dr_1 2 e^{-2r_1}, \nonumber
    \\
    \Gamma (C_{II}) = \exp{\left( - i \sigma_x \Sigma_{II}
    \right)}, \quad \Sigma_{II} = \int\limits_{S(C_{II})} dy dr_1
    2 e^{2r_1},
\end{eqnarray}
and $S(C_{I,II})$ are the regions in the planes
$(x,r_1)\left|_{\theta_1 = 0} \right.$ and
$(y,r_1)\left|_{\theta_1 = 0} \right.$ enclosed by the rectangular
loops $C_I$ and $C_{II}$ respectively. The sides of the rectangles
$C_I$ and $C_{II}$ are parallel to the coordinate axes. For the
loop $C_I$ these sides are given by the lines $r_1 = 0$, $x =
b_x$, $r_1 = d_x$ and $x= a_x$, where the length of the
rectangle's sides parallel to the $x$ axis is $l_x = b_x - a_x$.
In the Ref.~\cite{wePLA2} it was shown that
\begin{equation}\label{dx}
    d_x = - \frac{1}{2} \ln{\left( 1 - \frac{\pi}{4l_x} \right)},
    \quad l_x > \frac{\pi}{4} .
\end{equation}
In the same way the rectangle $C_{II}$ is composed of the lines
$r_1 = 0$, $y = b_y$, $r_1 = d_y$ and $y = a_y$,
where~\cite{wePLA2}:
\begin{equation}\label{dy}
    d_y = \frac{1}{2} \ln{\left( 1 + \frac{\pi}{2l_y} \right)},
    \quad l_y = b_y - a_y.
\end{equation}
We restrict ourselves by the consideration of the squeezing
control errors only. Moreover, we can neglect the fluctuations of
the squeezing control parameter when $r_1 = 0$. Thus to take into
account random squeezing control errors we have to replace $d_x$
by $d_x + \delta r_x (x)$ and $d_y$ by $d_y + \delta r_y (y)$,
where $\delta r_x (x)$ and $\delta r_y (y)$ are independent
Ornstein-Uhlenbeck stochastic processes. Making this substitution
into the Eqs.~(\ref{Gammas}) instead of the formulae~(\ref{HGG})
we obtain the following expression for the perturbed Hadamard
gate, see also~\cite{wePLA2}:
\begin{eqnarray}\label{PertHadamard}
    -i H = - \frac{1}{\sqrt{2}} \left( \cos{\alpha} - \sin{\alpha}
    \right) \left( \sin{\beta} + i \sigma_x \cos{\beta} \right) -
    \nonumber \\
    - \frac{i}{\sqrt{2}} \left( \cos{\alpha} + \sin{\alpha}
    \right) \left( \sigma_z \cos{\beta} - \sigma_y \sin{\beta}
    \right),
\end{eqnarray}
where
\begin{eqnarray}\label{albe}
    \alpha = e^{-2d_x} \int\limits_{a_x}^{b_x} dx \left( 1 - e^{-2\delta r_x}
    \right), \nonumber \\
    \beta = e^{2d_y} \int\limits_{a_y}^{b_y} dy \left( e^{2 \delta r_y} - 1
    \right).
\end{eqnarray}
Let the qubit initially to be in the pure state $\left| j \right>$
with $j$ equal to $0$ or $1$. Either for the fixed noise
realization or in the case of systematic errors the final qubit
state will be pure as well. However, it will differ from the
desired one. In the real experiment we do not follow the random
fluctuations of the control parameters (nevertheless {\it in
principle} we can do it). In this situation quantum mechanics
prescribes us to describe the final state of the system by the
density matrix and represent the state as a mixture of all
possible final states weighted with the probabilities of the
corresponding noise realizations. Following this strategy we find
the density matrix of the final state for a given noise
implementation and than average over the squeezing control
parameter fluctuations when the later are modeled by the two
independent Ornstein-Uhlenbeck stochastic processes.

Thus for the density operator $\tilde{\rho}_j \equiv H \left| j
\right> \left< j \right| H^\dag$ we obtain the following matrix
elements:
\begin{eqnarray}\label{tilderho}
    \left< j \right| \tilde{\rho}_j \left| j \right>  =
    \frac{1}{2} + \frac{1}{2} \cos{2 \gamma} \cos{2 \beta} ,
    \nonumber \\
    \left< nj \right| \tilde{\rho}_j \left| nj \right> =
    \frac{1}{2} - \frac{1}{2} \cos{2\gamma} \cos{2\beta},
    \nonumber \\
    \left< j \right| \tilde{\rho}_j \left| nj \right> = \left< nj
    \right| \tilde{\rho}_j \left| j \right>^\dag = \frac{i}{2}
    \sin{2 \beta} \cos{2\gamma} \nonumber \\ - \frac{1}{2} (-1)^j
    \sin{2\gamma}.
\end{eqnarray}
Here $\left| nj \right>$ means the state $\left| not \hspace{5pt}
j \right>$, for example, if $j=0$ than $nj = 1$, the introduced
parameter $\gamma$ is defined as $\gamma = \alpha - \pi /4$. From
the Eqs.~(\ref{tilderho}) it immediately follows that $tr
\tilde{\rho}_j = 1$ as it should be.

We assume that the noise $\delta r_x$ has variance
$\tilde{\sigma}_x$ and a lorentzian spectrum with the bandwidth
$\Gamma_x$. The fluctuations $\delta r_y$ have the variance
$\tilde{\sigma}_y$ and bandwidth $\Gamma_y$. Using the
Eqs.~(\ref{albe})-(\ref{tilderho}) and properties of
Ornstein-Uhlenbeck stochastic process  (see Ref.~\cite{Kaiser}) we
average the density matrix $\tilde{\rho}_j$ over the stochastic
fluctuations of the squeezing control parameters $\delta r_x$ and
$\delta r_y$. The averaged density matrix $\rho_j = \left<
\tilde{\rho}_j \right>$ has the following matrix elements:
\begin{eqnarray}\label{rhoj}
    \left< j \right| \rho_j \left| j \right> = \frac{1}{2} + 2
    e^{-2d_x} l_x \tilde{\sigma}_x , \nonumber \\
    \left< nj \right| \rho_j \left| nj \right> = \frac{1}{2} - 2
    e^{-2d_x} l_x \tilde{\sigma}_x , \nonumber \\
    \left< j \right| \rho_j \left| nj \right> = \left< nj \right|
    \rho_j \left| j \right>^\dag = \frac{(-1)^j}{2} \nonumber \\ - (-1)^j
    \frac{8\tilde{\sigma}_x}{\Gamma_x} e^{-4 d_x}
    \left[ l_x - \frac{1 - e^{- \Gamma_x l_x}}{\Gamma_x}
    \right].
\end{eqnarray}
Here we assumed that $\delta r_{x,y} \ll 1$ and restricted
ourselves by the first non-vanishing terms depending on $\delta
r_x$ or $\delta r_y$. The contribution of the stochastic control
errors made in the $(y,r_1) \left|_{\theta_1 = 0} \right.$ plane
can be neglected compared to the terms appeared due to the errors
made in the $(x, r_1) \left|_{\theta_1 = 0} \right.$ plane.

Now we find the fidelity of the non-ideal Hadamard gate. In the
case when there are no control errors ($\delta r_x = \delta r_y =
0$) the density matrix $\rho_{0j}$ of the final (pure) state has
the following matrix elements:
\begin{eqnarray}\label{rho0}
    \left< j \right| \rho_{0j} \left| j \right> = \left< nj
    \right| \rho_{0j} \left| nj \right> = \frac{1}{2} , \nonumber
    \\
    \left< j \right| \rho_{0j} \left| nj \right> = \left< nj
    \right| \rho_{0j} \left| j \right>^\dag  = \frac{(-1)^j}{2} .
\end{eqnarray}
The non-ideal Hadamard gate fidelity $F \equiv tr (\rho_{0j}
\rho_j)$ under the same assumptions as in the Eq.~(\ref{rhoj}) is
given by the expression
\begin{equation}\label{F}
    F = 1 - \frac{4 \tilde{\sigma}_x}{\Gamma_x l_x} \left( l_x \sqrt{2} - \frac{\pi}{2\sqrt{2}}
    \right)^2 \left[ 1 - \frac{1 - e^{- \Gamma_x l_x}}{\Gamma_x l_x}
    \right].
\end{equation}
In our previous work~\cite{wePLA2} the fidelity was defined as $f
= \sqrt{F}$. In the limit $(\Gamma_x l_x)^{-1} \to 0$, when the
fluctuations average out, from the Eq.~(\ref{F}) we obtain that
$1-f \sim \tilde{\sigma}_x^2$. It reproduces our previous
result~\cite{wePLA2} concerning the cancellation of the squeezing
control errors up to the fourth order on their magnitude (remind
that $\tilde{\sigma}_x$ has the order of $(\delta r_x)^2$).

Now we consider decoherence induced by the stochastic squeezing
control errors. In order to quantify decoherence strength we
exploit the purity of the final state. It is defined as the trace
of the squared density matrix. Purity equals to $1$ for pure
states and less than $1$ overwise. From the Eqs.~(\ref{tilderho})
it is easy to obtain that for a fixed noise realization or
equivalently in the case of systematic control errors the purity
$I_0 = tr \tilde{\rho}_j^2$ equals the unity. Thus the errors
equal for all qubits in the ensemble or for all consecutive
Hadamard gate implementations performed on a given qubit
(systematic control errors) do not lead to the decoherence and the
final state remains pure. Nevertheless, fidelity of the gate
implementation is less than unity in this case~\cite{wePLA1,
wePLA2}.

We use Eqs.~(\ref{rhoj}) to obtain the purity of the final state
in the case of the stochastic squeezing control errors. The result
can be expressed in a very simple form if we exploit the
expression~(\ref{F}) for the gate fidelity:
\begin{equation}\label{I}
    I = tr \rho_j^2 = \frac{1}{2} + \frac{1}{2} \left( 1 - 2 F
    \right)^2 \simeq 2F -1 .
\end{equation}
The last equality is hold if $(1 - F) \ll 1$. Thus we see that
stochastic squeezing control errors induce decoherence and lead
the final state to be a mixture of the pure states. Namely, if the
fidelity $F<1$ than the final state purity $I<1$.

In conclusion, in this Rapid Communication we considered optical
and ion trap HQC proposed the Refs.~\cite{OHQC} and \cite{TRIOHQC}
respectively. Regarding the particular implementation of Hadamard
gate we have studied decoherence induced by stochastic squeezing
control errors. Ornstein-Uhlenbeck stochastic process was
exploited to model random fluctuations of the squeezing control
parameter. We have analytically obtained the fidelity of the
non-ideal Hadamard gate and found the purity of the qubit's final
state. It was shown that the stochastic squeezing control errors
reduce the final state into a mixture of pure states and, thus,
induce decoherence. In the small errors limit a simple formulae
connecting the gate fidelity and the purity of the final state was
derived. In contradiction to the case of the stochastic control
errors systematic ones do not lead to decoherence and the final
state remains pure. Thus systematic control errors lead to wrong
output state only whereas stochastic control errors lead both to
wrong output and decoherence.

\end{document}